# Table-top all-attosecond transient absorption spectroscopy

Mikhail Volkov†, Evaldas Svirplys†, Stefanos Carlström†, Serguei Patchkovskii, Misha Yu. Ivanov, Marc J. J. Vrakking and Bernd Schütte*

Max-Born-Institut Berlin, Germany

†These authors contributed equally to this work.

*Corresponding author. Email: Bernd.Schuette@mbi-berlin.de

**Abstract:**

Attosecond transient absorption spectroscopy (ATAS) has emerged as a powerful technique within the field of attosecond science[1–4], combining extremely high temporal and excellent spectral resolution. So far, ATAS has been implemented in pump-probe experiments where an attosecond extreme-ultraviolet (XUV) pump or probe pulse was combined with a near-infrared (NIR) pulse in the femtosecond range, with the attosecond time resolution deriving from sub-cycle NIR-driven dynamics. Investigations of ultrafast electron dynamics in atoms, molecules and solids, with potential impact across physics, chemistry, and biology, would benefit significantly from the ability to perform *all-attosecond* transient absorption spectroscopy (AATAS)[5,6]. Here we demonstrate time-resolved AATAS using a table-top high-harmonic generation (HHG) source. The method is applied to investigate previously unresolved electronic coherences in Xe, revealing oscillatory valence hole motion with a 3-femtosecond period. In addition, systematic investigations of electron dynamics in Kr, Ar, and Ne are presented. Our work shows that, thanks to its broad bandwidth, high stability and easy accessibility, HHG is an ideal source for AATAS, offering the potential for replication in numerous laboratories.

**Introduction**

Electronic processes play a fundamental role in pivotal areas such as biology[7] and information technology[8]. The emergence of attosecond science[9–11] at the turn of this century has made it possible to investigate electron dynamics on extremely short timescales. To date, most attosecond pump-probe experiments have paired an attosecond pump or probe pulse with a femtosecond replica of the NIR driver pulse used to generate the attosecond pulses. This approach has resulted in an impressive body of work, subject to two significant limitations, namely (i) the strong NIR laser fields that are commonly used in two-color attosecond experiments are useful if one wishes to investigate field-driven effects, but hinder the observation of electron dynamics that is inherent to the system under investigation, and (ii) the typical ≥3 fs duration of the NIR pulses used makes it difficult to observe dynamics that occurs on 1-5 fs timescales. For example, while coherent hole motion could previously be observed in $Kr^+$ ($\tau$=6.3 fs) and in doubly- and triply-charged ions[12,13], it was not possible to observe coherent hole motion in $Xe^+$ ($\tau$=3.2 fs).

The most direct approach to studying electron dynamics on its natural timescale is attosecond-pump attosecond-probe spectroscopy (APAPS). APAPS requires the generation of intense attosecond pulses in order to ensure a sufficiently high excitation probability with the pump and/or probe pulse, which has limited its widespread and routine implementation. Thus far, a few proof-of-concept APAPS experiments have been demonstrated using table-top HHG sources[14–16]. These experiments relied on ion spectroscopy, which is straightforward to implement while providing indirect insight into the underlying electron dynamics. Following the first demonstration of attosecond pulses in Refs.[9,10], electron spectroscopy has proven very effective for studying attosecond electron dynamics within XUV-NIR experiments[17–19]. Recently, the Linac Coherent Light Source (LCLS) free-electron laser was used for first APAPS experiments employing electron spectroscopy[20,21]. The approach poses several challenges. For example, it is difficult in electron spectroscopy to achieve both the temporal and spectral resolution needed to measure attosecond or few-femtosecond electronic coherences, which arise from the excitation of coherent superpositions of quantum states whose energy differences determine their oscillation periods. In addition, space charge effects limit electron spectroscopy to dilute samples, meaning that APAPS based on charged particle detection is not suitable to study electron dynamics in liquids and solids.

Attosecond transient absorption spectroscopy (ATAS) overcomes these limitations, as it offers extremely high temporal resolution alongside excellent spectral resolution. Initially, ATAS was employed to observe coherent electron motion in atoms[1] and it has since proven highly effective in studying ultrafast electron dynamics in atoms, molecules and solids[2,3,22–24]. Two-color XUV+NIR ATAS is subject to the aforementioned limitations (i) and (ii). Therefore, the development of all-attosecond transient absorption spectroscopy (AATAS) is highly desirable. AATAS overcomes limitations (i) and (ii) and, as an additional benefit, permits the inclusion of element-specific excitation from valence and core levels in both the pump and the probe step of the experiment. A first AATAS experiment was recently carried out at LCLS, although the observation was limited to a single time delay[6].

Here we demonstrate time-resolved AATAS using attosecond extreme-ultraviolet (XUV) pulses as both the pump and probe. Utilizing a table-top HHG source, we achieve an inherently broad XUV bandwidth that is ideally suited for AATAS. Leveraging highly stable experimental conditions, previously unresolved coherent electron wavepacket motion in $Xe^+$ ions is observed in real time. Moreover, we observe an out-of-phase oscillation of inner- and outer-shell excitation absorption lines in $Ne^+$, which is rationalized with the help of an orbital picture of the coherent hole wavepacket motion.

**Table-top AATAS**

The peak power of attosecond pulses generated at FELs exceeds that of HHG by several orders of magnitude[21,25]. However, a more relevant consideration is to compare the photon fluence of the pump pulse, $F_\Upsilon$, in each laser shot, to the saturation fluence, $F_{\Upsilon,sat}=1/\sigma$, where $\sigma$ is the photoionization cross section. In our experiment $F_\Upsilon$=2.5x10$^{16}$ / cm$^2$, which is close to $F_{\Upsilon,sat}$=2.9x10$^{16}$ / cm$^2$ in the case of Ar and 4.6x10$^{16}$ /

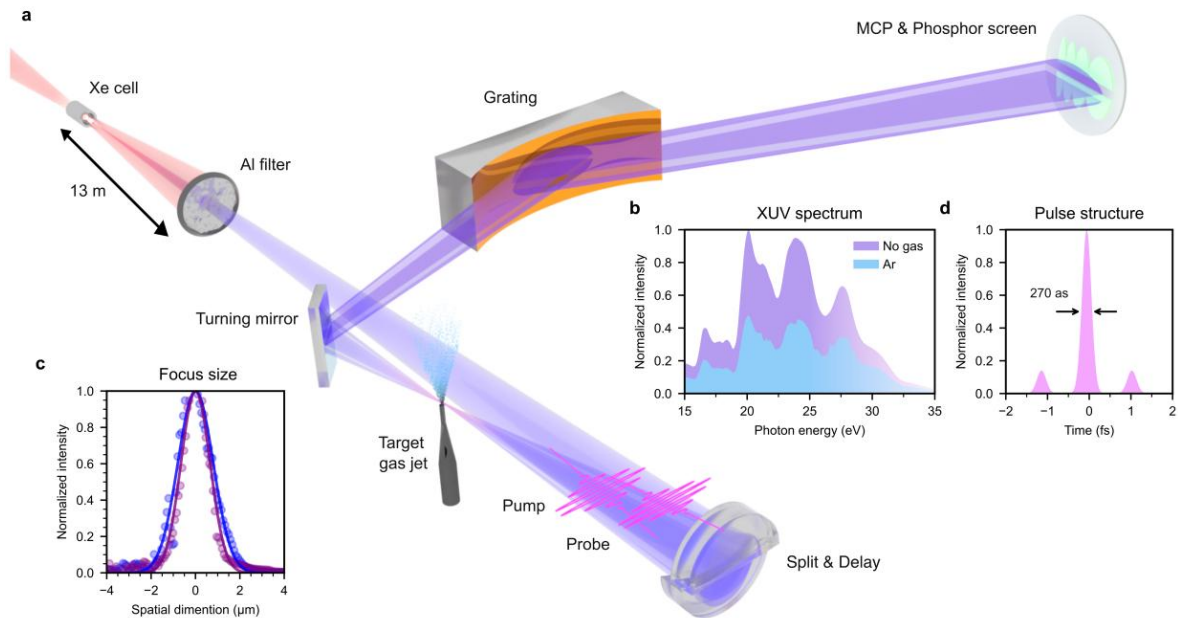

**Fig. 1. Experimental AATAS layout. a,** Schematic of the experiment. HHG is performed by focusing a near-infrared (NIR) beam into a gas cell filled with Xe. A telescope with an effective focal length of 3.5 m, consisting of a concave mirror (f = -1m) and a convex mirror (f = 0.75 m) is used to focus the pulses. Following attenuation of the NIR pulses using a 100-nm-thick aluminum (Al) filter, a split-and-delay unit (SDU) is employed 13 m downstream. The SDU consists of a nano-positioning stage and two spherical multilayer half-mirrors (focal length: 5 cm) with a reflectivity optimized in the 16-30 eV region. The SDU creates two replicas of the XUV pulse and allows for variation of their relative time delay. Absorption occurs in an effusive gas jet with a full width at half maximum of 46 μm. A turning mirror directs both the XUV pump and probe beams onto a grating, where they are diffracted onto a microchannel plate (MCP) / phosphor screen assembly. **b,** Spectra of the XUV pump pulse before (violet-shaded area) and after absorption in Ar (blue-shaded area). **c,** Focal profiles of the XUV pump beam in the vertical (blue) and horizontal (violet) directions, respectively, as measured using a knife-edge technique. The circles correspond to the measured transmission, which is fitted to a Gaussian profile (solid curves). The XUV pump beam waist radii in the vertical and horizontal directions are 1.3 μm and 1.6 μm, respectively. **d**, Temporal profile of the near-isolated attosecond pulses, as obtained by an autocorrelation measurement in Ar atoms[27].

cm$^2$ in the case of Xe (using $\sigma_{Ar}$=34.75 Mb and $\sigma_{Xe}$=21.53 Mb at 24 eV [26]). Further increasing the pump photon fluence would thus not significantly improve the signal-to-noise-ratio (SNR) of the data obtained in a pump-probe experiment. On the contrary, higher intensities would lead to excessive multiphoton absorption (see Supplementary Fig. 7), which is undesirable. It is noteworthy that, with a pump-laser induced differential absorbance on the order of a few percent, the transient signal changes observed in our experiments and in the experiments performed at LCLS[6] are comparable. In contrast to the pump pulse fluence, the probe pulse fluence does not affect the magnitude of the differential absorbance.

Once the pump photon fluence approaches the saturation fluence, further improvements of the SNR depend primarily on a reduction of and averaging over fluctuations of the experimental conditions. Key factors then are (i) the repetition rate at which the attosecond pulses are generated, (ii) the acquisition time over which data can be collected and (iii) the stability of the attosecond pulses. While both HHG and FEL sources used for APAPS may in future be operated at multi-kHz or even MHz repetition rates[16,21,28], HHG offers major advantages in terms of the available acquisition time and the stability. This directly expands the range of scientific questions that can be addressed.

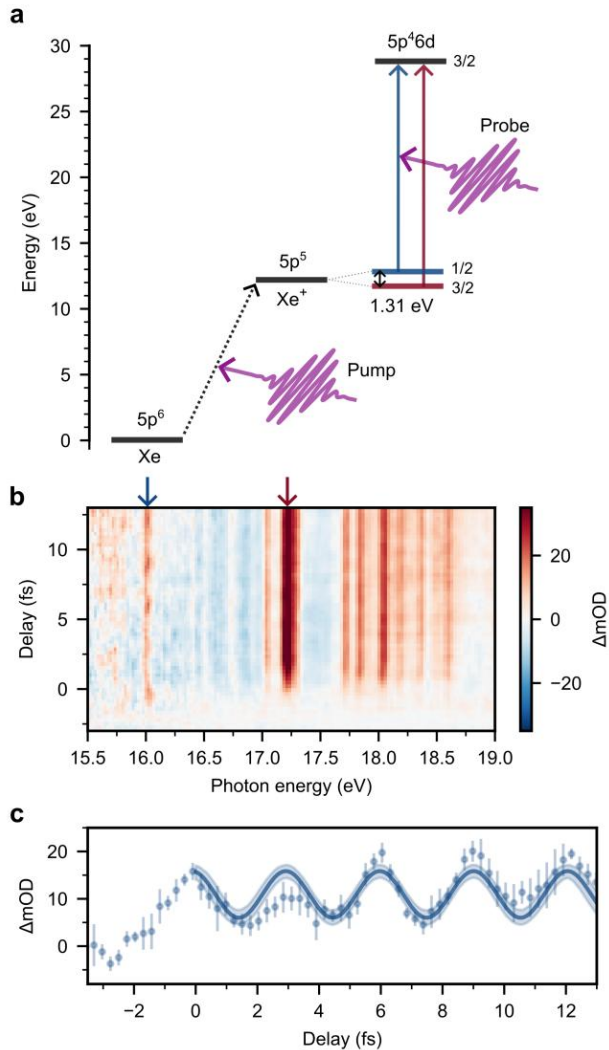

**Fig. 2. Observation of coherent electron motion in Xe$^+$. a,** The attosecond pump pulse generates Xe$^+$ ions in the ground and spin-orbit excited state, $5p^5_{3/2}$ and $5p^5_{1/2}$. The probe pulse resonantly excites the ions from these two spin-orbit states to higher-lying Rydberg states converging on Xe$^{2+}$. **b,** Transient changes of the optical density (OD) as a function of the photon energy and the time delay between the XUV pump and probe pulses. The attosecond pump pulse ionizes a fraction of the neutral Xe atoms in the focal volume, resulting in broadband reduction of the absorption of the probe pulse (shown in blue) when this pulse arrives after the pump pulse (positive time delays). In addition, narrow absorption features are visible, corresponding to resonant absorption in Xe$^+$. The two transitions that are indicated by the arrows correspond to excitation from each of the two spin-orbit states to the $5p^46d_{3/2}$ state. **c,** OD changes in the spectral region from 16.0 to 16.1 eV, corresponding to excitation from the $5p^5_{1/2}$ state to the $5p^46d_{3/2}$ state. The data were fitted by a sinusoidal function (solid curve), giving an oscillation period of 3.1±0.1 fs, as expected from the spin-orbit splitting in Xe$^+$.

In our laboratory we made systematic efforts to improve the signal-to-noise ratio in AATAS measurements. Near-infrared (NIR) pulses with an energy of 9 mJ and a duration of 37 fs were obtained from a commercial Ti:sapphire laser system operated at 1 kHz. This exceeds the repetition rate used in earlier APAPS experiments by 1-2 orders of magnitude[6,14,15,21]. To be able to generate near-isolated attosecond pulses, these NIR pulses were compressed to 3.7 fs using a three-stage compressor in a non-guided geometry, giving a pulse energy of 4.9 mJ. This approach resulted in an excellent single-shot NIR pulse energy stability of 0.17 % r.m.s.[27]. Further measures to improve the stability included a three-stage beam pointing stabilization system, temperature and humidity stabilization, as well as vibration damping.

To maximize the transient signal changes in our AATAS experiments, an HHG scheme was chosen that optimized the XUV intensity rather than the XUV pulse energy[29]. This was achieved using an 18-meter-long HHG beamline (Fig. 1**a**). The compressed NIR pulses were focused into an HHG cell using a telescope with an effective focal length of 3.5 m. The HHG efficiency was optimized by exploiting propagation effects of the NIR pulses in the HHG medium[30], resulting in an XUV pulse energy of about 200 nJ at the source[27]. By choosing a distance of 13 m between the HHG source and the XUV focusing mirrors with a focal length of 5 cm, a large de-magnification of the XUV source size was achieved. The pump and probe pulses were selected using two spherical XUV half-mirrors mounted on a split-and-delay unit. The obtained beam waist radii of the XUV pump pulses were 1.3 µm and 1.6 µm in the vertical and horizontal directions, respectively (Fig. 1**c**). The spatial chirp in the XUV focal plane was found to be negligible (see Supplementary Section 2). By taking into account the temporal attosecond pulse structure (Fig. 1**d**) and the pump pulse energy of 3.1 nJ, the XUV pump intensity was estimated as $2.8 \times 10^{14}$ W/cm$^2$, corresponding to the afore-mentioned photon fluence of $2.5 \times 10^{16}$ / cm$^2$. An XUV spectrometer was utilized to individually record the pump and the probe spectra with a spectral resolution between 30 and 70 meV. XUV spectra before and after absorption in Ar are displayed in Fig. 1**b**.

**Tracking coherent electron wavepacket motion**

The attosecond pump pulse ionized Xe, creating Xe$^+$ ions in a superposition of the spin-orbit ground and excited state (Fig. 2**a**), thereby initiating coherent hole motion. The energy splitting between the two spin-orbit states, $5p^5_{3/2}$ and $5p^5_{1/2}$, is 1.306 eV[31], corresponding to an oscillation period of 3.17 fs.

Fig. 2**b** presents an AATAS map recorded in Xe, illustrating the relative change in optical density (ΔOD) as a function of probe pulse photon energy and time delay. ΔOD was calculated using reference spectra at negative time delays, meaning that the probe pulse arrives before the pump pulse. For positive time delays, when the probe pulse arrives after the pump pulse, two distinct phenomena are visible: (i) Reduced absorption is observed across a broad photon energy range (16.1-17.0 eV) due to ionization of a fraction of the neutral Xe atoms by the pump pulse, thereby decreasing the density of neutral Xe gas in the interaction region. This bleaching effect, represented by the blue regions in the AATAS map, appears as a continuous feature across the displayed photon energy range; (ii) Enhanced absorption is evident at specific photon energies corresponding to absorption resonances in Xe$^+$ (red and white regions in the AATAS map). The two transitions that are indicated by arrows correspond to excitation from each of the two spin-orbit states to the $5p^4 6d_{3/2}$ state. The assignment of the other features is provided in Supplementary Section 8. Integration over the spectral range of 16.0–16.1 eV (Fig. 2**c**), corresponding to excitation from the $5p^5_{1/2}$ state to the $5p^4 6d_{3/2}$ state, reveals clear oscillations, which are attributed to coherent hole motion. The measured oscillation period of 3.1±0.1 fs reflects the spin-orbit splitting in Xe$^+$ [31].

The exceptional stability and robustness of our attosecond source facilitates systematic investigations of coherent hole motion. An AATAS map recorded in Ar is presented in Fig. 3**a**, showcasing the high quality of the recorded data, with a noise level estimated as 3 x 10$^{-4}$ OD. Assignments of the spectral lines can be found in Supplementary Section 8. Clear oscillations with a 23.3±0.1 fs period are observed, in agreement with the spin-orbit splitting of 0.177 eV in Ar$^+$ [32]. Integrated signals across three distinct energy regions are displayed in Fig. 3**b**, showing that oscillations at these different photon energies are out of phase.

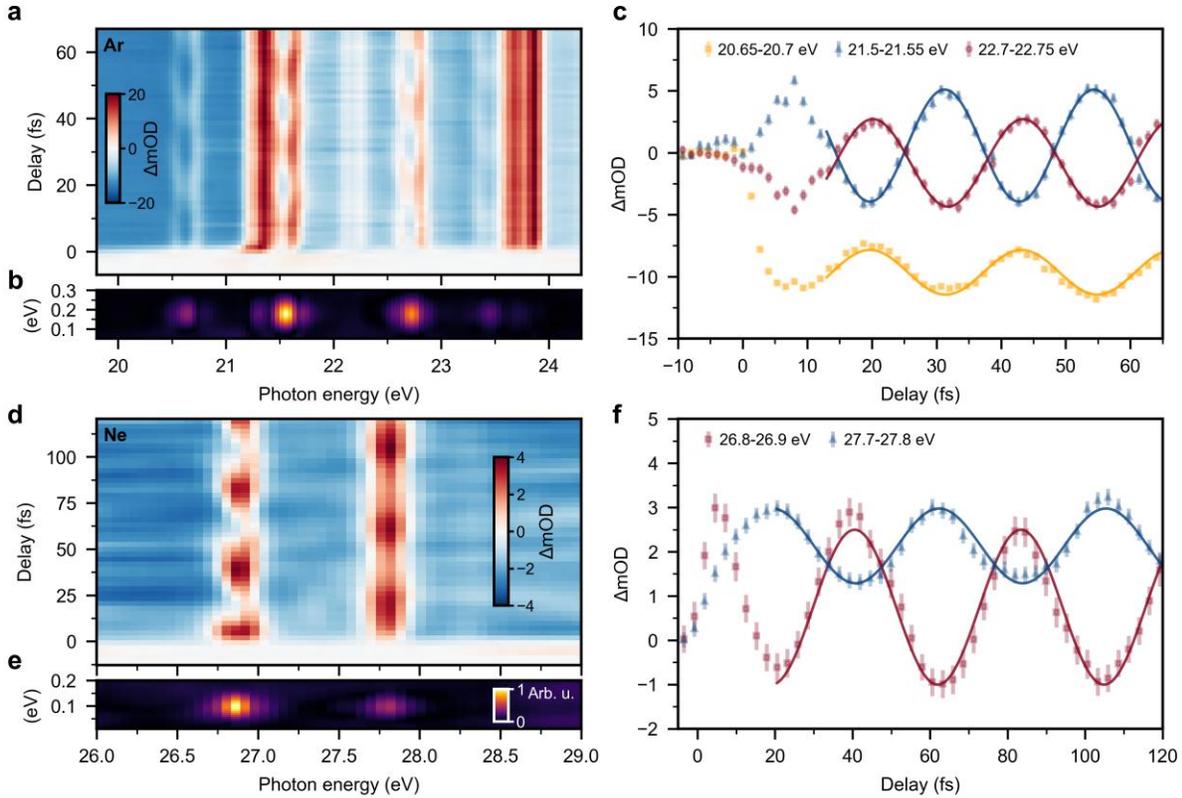

**Fig. 3. AATAS results in Ar and Ne. a**, Transient OD changes in Ar as a function of photon energy and time delay. **b**, Corresponding Fourier map with normalized amplitudes. The vertical axis corresponds to the oscillation frequency. **c**, OD variations at three distinct photon energies (data points) and corresponding fits (solid curves), showing oscillations with a 23.3±0.1 fs period. The phases of these oscillations differ across the spectrum. **d**, AATAS map in Ne and **e**, corresponding Fourier map. **f**, OD changes in Ne at two different photon energies, exhibiting out-of-phase oscillations with a 42.9±0.4 fs period, corresponding to the spin-orbit splitting of 0.097 eV in Ne$^+$. Assignments of the transitions can be found in Supplementary Section 8.

Performing AATAS in Ne enables access to the lowest excited states of Ne$^+$ at 26.8 and 26.9 eV[33], corresponding to excitation of a 2s inner valence electron into the 2p hole vacated by the ionization process, i.e. $2s^2 2p^5_{1/2} \rightarrow 2s^1 2p^6_{1/2}$ and $2s^2 2p^5_{3/2} \rightarrow 2s^1 2p^6_{1/2}$. Additional observed resonances around 27.8 eV correspond to the excitation of one of the remaining 2p electrons to a 3s orbital. For both types of observed resonances, the measured oscillation period of the coherent electron wavepacket motion is 42.9±0.4 fs (see Fig. 3**c**,**d**), in agreement with the spin-orbit splitting of 0.097 eV in Ne$^+$ [33]. Interestingly, like in the aforementioned case of Ar, out-of-phase behavior is observed across different photon energies (Fig. 3**d**) and will be discussed below. In addition to the results shown here, coherent electron wavepacket motion was also measured in Kr, see Supplementary Section 5.

**Theoretical calculations**

In order to interpret the out-of-phase oscillations observed in the Ne experiments, theoretical calculations were carried out simulating the electronic coherence resulting from the ionization process using time-dependent configuration interaction singles[34,35] by tracing out the photoelectron from the density matrix. The transient-absorption response was then computed at the single-atom level using perturbation theory[36].

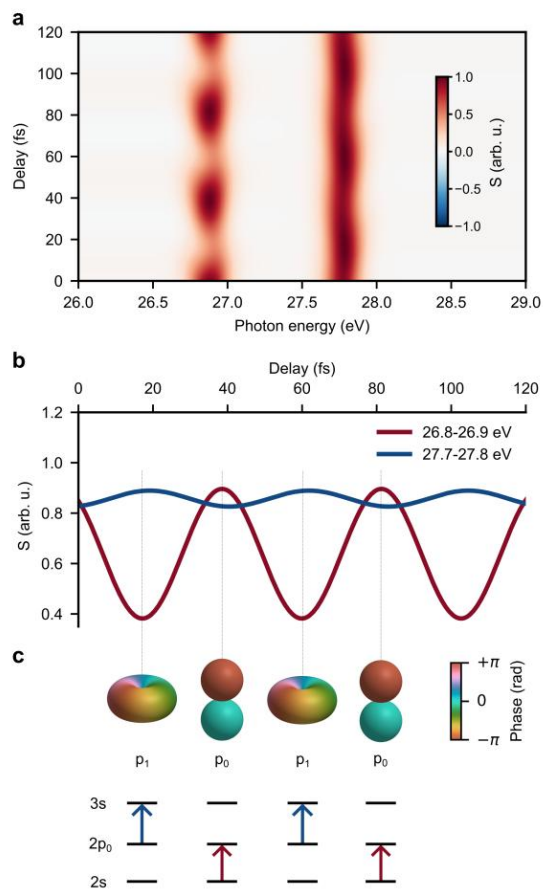

**Fig. 4: Simulation of AATAS in Ne. a**, Transient absorption map (*S*, arbitrary units), calculated using perturbation theory (See Supplementary Section 9 for more details). The single-atom response has been convoluted with a Gaussian of 70 meV bandwidth to simulate the instrument response. **b**, Lineouts of two absorption features at photon energy intervals of 26.8-26.9 eV (red curve) and at 27.7-27.8 eV (blue curve), showing out-of-phase oscillations. **c**, The hole density oscillates between $p_0$ and $p_1$ character as a function of the pump–probe delay (see also Supplementary Fig. 9). The shape of this hole determines which electron transition (2s→$2p_0$ or $2p_0$→3s) is more likely after excitation by a delayed probe pulse. When the hole is of **$p_1$** character, an electron transition from an occupied $2p_0$ orbital to an unoccupied 3s orbital is more likely (indicated by a blue arrow). In contrast, when the hole is of **$p_0$** character, electron transfer from an occupied 2s orbital to an unoccupied $2p_0$ orbital is maximized (red arrow), producing the out-of-phase oscillations of the two absorption features in the AATAS map.

The dipole moments were calculated using DBSR[37], whereas the energies were taken as the experimental ones[38]. More details regarding the calculations are provided in Supplementary Section 9.

According to the simulations, two oscillations that occur near 26.85 eV (corresponding to the transitions from the two spin-orbit states of $Ne^+$ to $Ne^+(2s2p^6)_{J=1/2}$ are in phase with each other, leading to an oscillation with high contrast (see Fig. 4**a**). In the spectral range near 27.75 eV, four closely spaced transitions contribute, corresponding to transitions of the two spin-orbit states of $Ne^+$ to $Ne^+(2s^22p^43s)_{J=1/2,3/2}$ [39]. Two of the transitions oscillate in phase with the 26.85 eV oscillations, and two of them oscillate out of phase with these oscillations. Taking into account the spectrometer resolution, a high-contrast oscillation is visible in the 26.8-26.9 eV interval, along with an out-of-phase, lower contrast oscillation in the 27.7-27.8 eV interval (Fig. 4**b**), in agreement with the experimental results (see Fig. 3**d**).

To rationalize the out-of-phase behavior, the time-dependent hole density obtained from the calculations needs to be considered. In *jj*-coupling, and for $m=1/2$, the two spin-orbit states are given, respectively, as $\left|J=\tfrac{1}{2}, m=\tfrac{1}{2}\right\rangle = -\sqrt{\tfrac{1}{3}}\, 2p_0\alpha + \sqrt{\tfrac{2}{3}}\, 2p_1\beta$ and $\left|J=\tfrac{3}{2}, m=\tfrac{1}{2}\right\rangle = \sqrt{\tfrac{2}{3}}\, 2p_0\alpha + \sqrt{\tfrac{1}{3}}\, 2p_1\beta$, where $2p_0$ and $2p_1$ describe $\boldsymbol{l} = \boldsymbol{1}$ holes in the $n=2$ shell with $\boldsymbol{m_l = 0}$ and $\boldsymbol{m_l = 1}$, respectively, and $\boldsymbol{\alpha}$ and $\boldsymbol{\beta}$ are spin labels. Upon ionization, a partially coherent wavepacket is created in the ion with an initial excess in the population of $\mathbf{p_0}$. This wavepacket will oscillate with the characteristic time period given by the spin–orbit splitting. The shape of the hole will oscillate between a dumbbell-like shape (indicated by $p_0$ in Fig. 4**c**) and a doughnut-like shape (indicated by $p_1$), with connecting intermediate shapes shown in Supplementary Figure 9.

For the probe parameters of the experiment, three states are qualitatively relevant: the spin–orbit split ground state of the ion $2s^2 2p^5\ ^2P_{J=3/2,\,1/2}$ ($E_{3/2} = 0$ eV, $E_{1/2} = 0.097$ eV), the valence-excited state $2s 2p^6\ ^2S_{1/2}$ ($E = 26.910$ eV), and the Rydberg-excited state $2s^2 2p^4 3s\ ^2P_{1/2}$ ($E = 27.859$ eV). In a particle–hole picture of the *LS* coupling scheme, the valence-excited state predominantly corresponds to excitation from the 2s inner-valence orbital to the $p_0$ outer-valence orbital. Similarly, the Rydberg-excited state is formed predominantly from excitation of the $p_0$ orbital to the 3s orbital.

Transfer of an electron from an occupied 2s orbital to an unoccupied $p_0$ orbital is maximized when the hole is localized on the $p_0$ orbital. Given the initial excess population of $p_0$, the oscillation around 26.85 eV therefore starts with maximum absorption near the time overlap between the pump and the probe pulses. Conversely, excitation from the $p_0$ orbital to an unoccupied 3s orbital is maximized when the population in the $p_0$ orbital is maximal, and when the hole is localized on the doughnut ($p_1$). For this reason, the absorption at 27.75 eV starts near a minimum and gains in strength when the population of the $p_0$ orbital is decreased and that of the $p_1$ orbital is increased.

**Discussion**

In summary, we have demonstrated time-resolved all-attosecond transient absorption spectroscopy (AATAS) using XUV pulses as both pump and probe. The remarkable stability of our attosecond light source allowed us to obtain high-quality data. In combination with its broad bandwidth and wide accessibility, HHG is shown to be an ideal source for AATAS. Table-top AATAS may be extended to higher photon energies in the XUV and soft X-ray ranges, enabling core-level spectroscopy in atoms and molecules. Sufficiently high intensities have already been demonstrated around 100 eV[40]. At higher photon energies, the use of higher repetition rates and longer acquisition times can help compensate for the lower signals resulting from lower intensities and smaller cross sections.

The presented approach can readily be replicated, as similar beamline infrastructures are available in laboratories worldwide[14,15,40–46]. Additionally, we have previously shown that intense XUV pulses can be generated using a compact, two-meter-long setup[47], which recently enabled the first all-attosecond pump-probe experiment at a kHz repetition rate[16]. These advancements suggest that traditional attosecond setups can evolve into systems fully capable of performing AATAS.

Our method offers groundbreaking opportunities for exploring ultrafast electron dynamics in molecules and solids. Unlike traditional attosecond pump-probe experiments that rely on strong NIR fields, our approach eliminates the need for such fields. This enables a significantly simpler analysis and more straightforward interpretation of the observed electron dynamics. Moreover, in the field of attochemistry, our approach may be used to investigate coupled electronic and nuclear dynamics evolving during the passage through conical intersections[22,48].

The noise in AATAS measurements can be further suppressed in the future by implementing advanced referencing methods[49–52]. Such improvements could maintain the same SNR using significantly lower XUV intensities. This is especially beneficial for the investigation of solids, as it minimizes the risk of sample degradation. Since AATAS enables specific excitation from both valence and core levels in both the pump and the probe steps, it opens the door to studying complex electron dynamics in solids. Considering that a significant number of elements have inner-shell binding energies in the spectral region of our experiment[53], core-hole decay processes and charge transfer dynamics[54] could be investigated directly in the time domain.

**Acknowledgments:** We thank Melanie Krause, Christoph Reiter and Roman Peslin for technical support.

**Funding:**
Deutsche Forschungsgemeinschaft, project numbers 456137830 and 471478110

**Author contributions:**
B.S. and M.J.J.V. conceived the experiment. B.S., M.V. and E.S. developed the experimental setup and performed the measurements. E.S. and M.V. analyzed the experimental data. S.C. and S.P. performed the theoretical calculations. M.Y.I. supervised the theoretical work. All authors contributed to writing the manuscript.

**Competing interests:** The authors declare that they have no competing interests.

**Data and materials availability:** The data that support the findings of this study are available from the corresponding author upon reasonable request.